\documentclass[12pt,preprint]{aastex}

\usepackage{amsmath}

\newcommand{\etal}{{\em et al.}}
\newcommand{\Hii} {[H{\sc ii}]}
\newcommand{\sii} {[S{\sc ii}]}
\newcommand{\Ha}  {H$\alpha$}
\newcommand{\msol}{\hbox{M$_{\odot}$}}
\newcommand{\kms} {\hbox{km~s$^{-1}$}}
\newcommand{\vlsr}{\hbox{$v_{LSR}$}}
\newcommand{\tcoa}{\hbox{$^{13}$CO$_{1-0}$}}
\newcommand{\Tcoa}{\hbox{$^{12}$CO$_{1-0}$}}
\newcommand{\Tcob}{\hbox{$^{12}$CO$_{2-1}$}}
\newcommand{\mum} {$\mu$m}

\slugcomment{Accepted for publication in The Astrophysical Journal, April 2006 (vol 641)}
\shorttitle{Star Formation in L1551}
\shortauthors{Moriarty-Schieven \etal}

\begin{document}

\title{Multi-Generational Star Formation in L1551}

\author{Gerald H. Moriarty-Schieven}
\affil{National Research Council of Canada, Joint Astronomy Centre, 
660 N. A'ohoku Pl., Hilo, HI  96720}
\email{g.schieven@jach.hawaii.edu}

\author{Doug Johnstone}
\affil{National Research Council of Canada, Herzberg Institute of
Astrophysics, 5071 West Saanich Road, Victoria, BC V9E 2E7, Canada} 
\affil{Department of Physics and Astronomy, Univerisity of Victoria, 
Victoria, BC, V8P 1A1, Canada}
\email{doug.johnstone@nrc-crnc.gc.ca}

\author{John Bally}
\affil{Department of Astrophysical and Planetary Sciences,\\
       Center for Astrophysics and Space Astronomy,\\
       Campus Box 389, University of Colorado, Boulder CO 80309} 
\email{bally@casa.colorado.edu}

\and

\author{Tim Jenness}
\affil{Joint Astronomy Centre, 
660 N. A'ohoku Pl., Hilo, HI  96720}
\email{t.jenness@jach.hawaii.edu}

\clearpage

\newpage

\begin{abstract}

The L1551 molecular cloud, unlike most of the Taurus 
Molecular Complex, is undergoing a long and sustained period of 
relatively high efficiency star formation. 
It contains two small clusters of Class 0 and I protostars, as well 
as a halo of more evolved Class II and III YSOs, indicating a current 
and at least one past burst of star formation.  We present here  
new, sensitive maps of 850 and 450 \mum\  dust emission covering most of the 
L1551 cloud, new CO J=2-1 data 
of the molecular cloud, and a new, deep, optical image of \sii\ emission  
(6730\AA).  We have detected all of the previously known Class 0 and I 
YSOs in L1551, and no new ones.  Compact sub-millimetre emitters are 
concentrated in two sub-clusters:  IRS5 and L1551NE, and the HL~Tauri 
group.  Both stellar groups show significant extended emission and outflow/jet 
activity.  A jet, terminating at HH~265 and with a very weak 
associated molecular outflow, may originate from LkH$\alpha$~358, or 
from a binary companion to another member of the HL~Tauri group.   
Several Herbig Haro 
objects associated with IRS5/NE were  
clearly detected in the sub-mm, as were faint ridges of emission 
tracing outflow cavity walls.  We confirm a large-scale molecular 
outflow originating from NE parallel to that from IRS5, and suggest 
that the ``hollow shell'' morphology is more likely due to two 
interacting outflows.  The origin of the E-W flow east of HH~102 is 
undetermined.  We confirm the presence of a prestellar core (L1551-MC) 
of mass 2-3 M$_{\odot}$ north-west of IRS5.  The next generation 
cluster may be forming in this core.  The L1551 cloud appears cometary 
in morphology, and appears to be illuminated and eroded from the 
direction of Orion, perhaps explaining the multiple 
episodes of star formation in this cloud. 

The full paper (including figures) can be downloaded at http://www.jach.hawaii.edu/$\sim$gms/l1551/l1551-apj641.pdf, or viewed at http://www.jach.hawaii.edu/$\sim$gms/l1551/ .
 
\end{abstract}

\keywords{stars:formation; ISM:individual(LDN1551); ISM:jets and
outflows; submillimeter}

\clearpage

\section{Introduction}


In stark contrast to most of the Taurus-Auriga star formation complex,
which is characterized by very low efficiency formation of isolated
stars (Palla \& Stahler 2002), the L1551 molecular cloud has sustained
a long and continuing period of star formation of relatively high
efficiency. 
The cloud is surrounded by a halo of at least 30 classical and weak T Tauri
stars (cTTs, wTTs) and proto-brown dwarfs 
(Cudworth \& Herbig 1979; Jones \& Herbig 1979; Feigelson \& DeCampli
1981; Feigelson \& Kriss 1983; Wood \etal\ 1984; Feigelson \etal\ 1987;
Gomez \etal\ 1992; Carkner \etal\ 1996; Brice{\~n}o, {\em et al.} 1998;
Brice{\~n}o, Stauffer \& Kirkpatrick 2002; Favata \etal\ 2003;
G{\aa}lfalk \etal\ 2004), indicating star formation 
activity over the past few million
years.
Embedded within the cloud 
are two known groups of active star
formation, one comprised of IRS5 and L1551-NE (hereafter NE), and the
other the HL~Tau group, located $\sim$5\arcmin\ north of IRS5 and consisting
of HL~Tau, XZ~Tau, LkH$\alpha$ 358, and HH~30*.   (In the following discussion,
HH~30 refers to the Herbig-Haro object(s), while HH~30* (also known as
V1213 Tau) refers to the driving star.)
The question
remains, will this sustained star formation continue beyond the
present epoch, and what is the reason for this sustained activity?

LDN 1551 (Lynds 1962), a modest-sized ($M \approx$ 50 \msol, diameter
$D \sim$ 1 pc) 
(Moriarty-Schieven \& Snell 1988) dark cloud, lies directly behind
the Hyades star cluster in Taurus.
Its distance 
was recently constrained using Hipparchos observations by
Bertout, Robichon \& Arenou (1999), who determined that the average
distance to the Taurus-Auriga complex was 139$^{+10}_{-9}$ pc, but
found the distance to T Tauri (which is much nearer to L1551 than the
rest of the Tau-Aur complex) to be 168$^{+12}_{-28}$ pc.
Extended optical emission from L1551 was first
discovered on photographic plates and cataloged as the diffuse
\Hii\ region S239 (Sharpless 1959).
This emission was later
found to exhibit characteristics of a Herbig-Haro object (shock
excited optical emission powered by outflows from young stars)
and re-categorized as HH~102 (Strom, Grasdalen, \& Strom 1974).
Three bright and compact HH objects, HH~28, HH~29, and HH~30, were 
also found in the region (Herbig 1974).  
They were at first
mis-identified as high velocity stars due to their high proper
motions (Luyten 1963, 1971).
Recent deep images
of the L1551 cloud have led to the discovery of additional
Herbig-Haro objects (Graham \& Heyer 1990; L\'{o}pez \etal\ 1995;
Riera \etal\ 1995).

New aspects of the L1551 star forming region continue to be uncovered as
deep optical images, high angular resolution radio continuum maps, and 
other studies are obtained.  High angular resolution radio continuum 
maps have shown that IRS~5, an {\em FU Ori} type Class I protostar
(Sandell \& Weintraub 2001), is a binary (Rodriguez \etal\ 1986, 2003a)
with a separation of 
about 50 AU (Looney, Mundy \& Welch 1997; Rodriguez \etal\ 1998),
total mass $\sim$1.2$_{\odot}$, and period $\sim$260yr (Rodr{\'i}guez
\etal\ 2003a). 
Each star in this system drives a distinct stellar jet 
(Snell \etal\ 1985; Fridlund \& Liseau 1998; Rodr{\'i}guez \etal\ 2003b)
with orientations that are misaligned with the CO outflow lobes,
discovered by Snell, Loren \& Plambeck (1980), 
and thought to be powered by IRS~5.  Deep optical imaging, proper 
motion studies, and spectroscopy have provided evidence that 
the nearby YSO L1551~NE, a Class 0 protostar (Moriarty-Schieven,
Butner \& Wannier
1995) discovered by Emerson \etal\ (1984), located
about 1\arcmin\ to the east of 
IRS~5, may power the bright Herbig-Haro object HH~29 and
possibly HH~28 (Devine, Reipurth, \& Bally 2000; Hartigan et al. 2000)
along with
a highly collimated 1.5 parsec-long chain of fainter
Herbig-Haro objects that are superimposed on the main CO lobes
of the L1551 outflow complex.  The NE source is also a multiple system
(Rodriguez, Anglada \& Raga 1995; Moriarty-Schieven \etal\ 2000), and
possesses a molecular outflow 
(Moriarty-Schieven, Butner \& Wannier 1995), although the outflow is
aligned with  
and hence confused by that of IRS5.  Associated with IRS5/NE is
another low-velocity molecular outflow, the L1551 E-W flow
(Moriarty-Schieven \& Wannier 1991; Pound \& Bally 1991), which
extends $\sim$30\arcmin nearly due west of IRS5/NE, but whose origin is unknown.

Near HL~Tauri is a compact group of four or five YSOs,  of which all
but one have
outflows and/or jets (Mundt, Buhrke \& Ray 1988).  
Deep images obtained through
narrow-band interference filters transmitting the light of \Ha\
and \sii\ show that the nearby sources HL and XZ~Tau and the
source of the HH~30 stellar jet power optically visible
Herbig-Haro flows which also criss-cross the IRS~5 outflow
lobes (Devine, Reipurth, \& Bally 1999).  Several molecular outflows
have also been detected toward this group (Monin, Pudritz \& Lazareff 1996).

Most of the surveys for dense cores in the L1551 cloud have
concentrated on the IRS5 region (e.g. Moriarty-Schieven \etal\ 1987 for
CS, Chandler \& Richer 2000 for dust continuum), while Onishi \etal\
(2002) neglected L1551 in their systematic H$^{13}$CO$^{+}$ survey of cores
in Taurus.  In this work we present 850 \mum\  and 450 \mum\  dust
continuum images of a region covering most of the L1551 cloud.  In
addition, we present new CO J=2-1 maps of the cloud, in order to
search for new outflows and disentangle those already known.

The full paper (including figures) can be downloaded at

\noindent http://www.jach.hawaii.edu/$\sim$gms/l1551/l1551-apj641.pdf,

\noindent or viewed
at http://www.jach.hawaii.edu/$\sim$gms/l1551/ . 

\section{Observations and Data Reduction}

\subsection{850/450\mum\  Continuum Data}

850 \mum\  and 450 \mum\  continuum data were obtained using the SCUBA
bolometer array (Holland et al. 1999) mounted on the James Clerk
Maxwell Telescope (JCMT)\footnote{The JCMT 
is operated by the Royal Observatory Edinburgh on
behalf of the Particle Physics and Astronomy Research Council
of the United Kingdom, the Netherlands
Organization for Scientific Research, and the National Research Council of
Canada.} located near the summit of Mauna Kea, Hawaii.  The
20\arcmin$\times$20\arcmin\ region centered on IRS5 was mapped on 12-13 January
2001 using a scan-mapping technique.  Sky transparency for these
nights was moderately good to good, with $\tau_{225GHz} \sim$ 0.06 on
12 Jan and $\sim$ 0.075 on 13 January.  In addition to these data, the
SCUBA archive was ``mined'' for all observations (pointing,
calibration, mapping, etc.) taken toward this
region between 1997 June and 2002 June (approximately 300 observations
in all), and included in the final 
reduction.

\begin{figure}
 \epsscale{1.0}
 \includegraphics{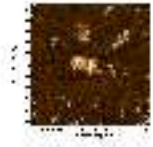}
 \vspace*{1.25in}
 \caption{850\mum\ image of the L1551 molecular cloud revealing the HL
Tau region  
(north), the IRS~5/NE region (center), the L1551 outflows, (left and
right), and 
the L1551-MC region (upper right).  Contour intervals are at 0.035,
0.07, 0.14, 0.28, 0.56, 1.12 and 2.24 Jy beam$^{-1}$.
\label{850um}} 
\end{figure}

\begin{figure}
 \epsscale{1.0}
 \includegraphics{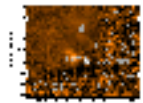}
 \vspace*{1.25in}
 \caption{450\mum\ image of L1551 molecular cloud toward HL~Tau and 
IRS5/NE.  Contour intervals are at 0.15, 0.3, 0.6, 1.2, 2.4 and 4.8 
Jy beam$^{-1}$.
\label{450um}} 
\end{figure}

The initial data reduction (switch removal, extinction correction,
despiking, etc., but not sky subtraction which was corrected later)
were performed using the {\em oracdr} and {\em surf} 
data reduction packages (Jenness et al., 2002).  The data were flux
calibrated 
using the archival values derived by Jenness \etal\ (2002).  Images were
reconstructed using a matrix inversion method described by Johnstone
et al. (2000).  The reconstructed images are, however, still subject
to large-scale ``features'' which are artifacts of the reconstruction
and may not be real.  We removed these large-scale structures ($>
\sim$ few arcmin) with unsharp masking.  The final 850 \mum\  image is
shown in Figure \ref{850um}.  The 450 \mum\  image is shown in Figure
\ref{450um}.

\begin{figure}
 \epsscale{1.0}
 \includegraphics{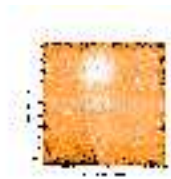}
 \vspace*{1.25in}
 \caption{RMS variance for the 850\mum\ image. Note that the uncertainty is 
much less in regions where multiple measurements have been obtained, primarily 
toward the HL~Tau region, where the variance reaches a minimum of
0.0017 Jy beam$^{-1}$.  The white contour is at 0.010 Jy beam$^{-1}$,
black contours are at 0.012, 0.014 and 0.016 Jy beam$^{-1}$, and grey
contours run from 0.008 to 0.002 Jy beam$^{-1}$ at intervals of 0.002
Jy beam$^{-1}$. 
Apparent variance peaks toward IRS5, NE and
HL~Tau are artifacts of the reconstruction process.  
\label{850um-var}} 
\end{figure}

Because of archive data mining, some portions of the region,
particularly toward HL~Tau (which is a SCUBA secondary flux calibrator),
have been observed multiple times, and hence the RMS noise of the final 
image is significantly lower in these locations.  In 
Figure \ref{850um-var} we show a greyscale image
of the variance across the image, with contours at 10 mJy beam$^{-1}$
intervals.  (Note that the variance towards HL~Tau, IRS 5 and NE has
been skewed by the brightness of the sources.)  In the
$\sim$5\arcmin$\times$5\arcmin\ region centered on HL~Tau, the RMS is $\sim$2
mJy beam$^{-1}$, towards IRS5/NE the RMS is $\sim$6 mJy beam$^{-1}$, and throughout
much of the rest of the image the variance is $\sim$10 mJy beam$^{-1}$.  At
450 \mum\  the RMS in the region near HL~Tau is $\sim$15 mJy beam$^{-1}$,
but $\sim$70 mJy beam$^{-1}$ through most of the image.

\subsection{\sii\ }

Images were obtained on the nights of December 15, 17, and 19, 2001 
with the NOAO MOSAIC2 Camera at the f/3.1 prime focus of the 4 meter
Blanco telescope at the Cerro Tololo Interamerican Observatory.  The MOSAIC 
camera is 8192$\times$8192 pixels (consisting of eight 2048$\times$4096 pixel
CCD chips) with a pixel scale of 0.26'' pixel$^{-1}$ and a field of
view 35.4' on a side.  We used a narrow-band filter centered on
6730\AA\ and 6596\AA\
with a FWHM of 80\AA\ for our \sii\ and H$\alpha$ 
observations respectively.  Continuum (centered on 7732\AA\
with a FWHM of 1546\AA) images were also obtained, but will be presented in
another paper.

In order to increase our observing efficiency, we opted not to take a
set of five images in the standard MOSDITHER pattern which is normally
used to eliminate cosmic rays and the gaps between the individual
chips in the MOSAIC camera.  Instead, we took a single image in each
filter for each pointing.  As a result of this observing strategy,
most of our images still contain chip gaps, resulting in a loss of
approximately 3\% of the area coverage at each pointing. However, using
this strategy, we were able to effectively cover a 7.5 square degree
area in our allotted three nights.  

Images were overscanned, trimmed, bias subtracted, and flat fielded
(using dome flats) in the standard manner using the MSCRED package in
IRAF.  Cosmic rays were removed using
CRNEBULA\footnotemark.  Because we used single exposures in much of
the survey, the images may contain a small number of cosmic rays which
were not eliminated by the CRNEBULA task, however we can safely
distinguish the cosmic rays from objects by their morphology: cosmic
rays have sharp edges and are often only one pixel in extent.

\footnotetext{The CRNEBULA task removes cosmic rays from a region with
fine nebular structure which can be misidentified by more traditional
cosmic ray rejection routines.  The routine uses box and ring median
filters to distinguish fine nebular structure from cosmic rays.  For a
detailed discussion of how this procedure works, see the IRAF CRNEBULA
help page (available at
http://iraf.noao.edu/scripts/irafhelp?crnebula).}

We used the procedures MSCCMATCH, MSCIMAGE, and MSCIMATCH to
remove relative distortions, generate a single extension FITS image,
and match the sky background between images.  For the pointings for
which we obtained multiple images in each filter in a dither pattern,
we used the MSCSTACK procedure to combine all images in each filter 
into a single image which eliminates the gaps between CCD chips.

\begin{figure}
 \epsscale{1.0}
 \includegraphics{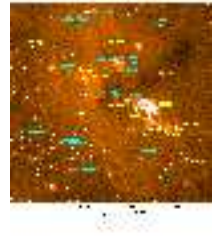}
 \vspace*{1.25in}
 \caption{\sii\ image of L1551 cloud.  Known PMS and YSOs are labeled
along
with selected Herbig-Haro objects. Note the dark arc to the north-west
of IRS5.
\label{sii}} 
\end{figure}

A portion of the \sii\ image, covering the L1551 cloud, is shown in
Figure \ref{sii}.

\subsection{CO $J=2-1$ Emission}

CO J=2-1 observations were obtained at the (former) 
National Radio Astronomy
Observatory (NRAO) 12 meter diameter Cassegrain radio telescope
located on Kitt Peak Arizona in January and February of 1997. 
The telescope has a pointing accuracy of 5\arcsec, an aperture
efficiency of 32\%, and main beam efficiency of 44\% at
230~GHz. At 230~GHz, the beamsize is 27\arcsec.  The 200 to
265~GHz SIS receiver had a single-sideband receiver noise
temperature of about 200~K providing a system temperature
around 500~K.  This dual channel receiver acquires data in
two orthogonal polarization states simultaneously. The system
back-end consisted of a 768~channel hybrid spectrometer with
each channel delivering a bandwidth of 195.3~kHz for a total
bandwidth of about 150~MHz.  At 230~GHz this corresponds to
0.25~\kms\ per channel and an overall bandwidth of 192~\kms. 

Spectra were acquired using on-the-fly (OTF) mapping, wherein
the telescope is rastered rapidly across the sky while data and
antenna position information are recorded at a 10~Hz rate
(Mangum 1997). The telescope was rastered across 30\arcmin\
rows. After each row, the telescope was moved 6.9\arcsec\
perpendicular to the row and scanned in the opposite direction.
Following the completion of each pair of rows, the telescope
was position switched to an emission-free reference position
and a calibration observation (ambient temperature load and sky
emission) was obtained. The reference position was located at
$\alpha$(2000) = 4$^h$32$^m$57$^s$.87, $\delta$ (2000) =
17\arcdeg 52\arcmin 58\arcsec, which was found to be
emission-free to 0.1~K at all velocities.  The telescope
pointing was checked several times during each observing
session and the carbon star IRC+10216 was used as an absolute
intensity calibration source.  

The telescope raster orientation was chosen to be either
aligned with or orthogonal to a position angle of 60\arcdeg\
which is close to the orientation of the IRS~5 outflow. Each
field was observed many times with alternating orthogonal
raster orientations. The resulting map covers a 30\arcmin\ by
52\arcmin\ field with a major axis tilted at $PA$ = 60\arcdeg.
There are substantial horizontal and vertical stripes in the
raw images along the principle directions of the OTF scanning
(at $PA$ = 60\arcdeg\ and $PA$ = 150\arcdeg). This artifact is
produced by high order baseline ripples and calibration
differences between receivers. The striping was partially
eliminated by cross-calibration of orthogonal scans (`basket
weaving') and by fitting higher order baselines to the data. 

The entire data set was reduced using 
the AIPS data reduction package. The reduced spectra
were re-sampled onto a uniform 10\arcsec\ grid and first-order
baselines were subtracted.  The resulting rms noise per
195.3~kHz channel is about 0.1~K at each grid point. All
further reductions and analyses were conducted with the IDL and
COMB data analysis packages. 

\begin{figure}
 \epsscale{1.0}
 \includegraphics{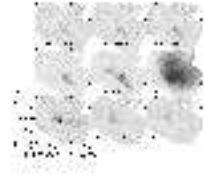}
 \vspace*{1.25in}
 \caption{Channel maps of CO J=2-1 emission.
\label{co-chan}} 
\end{figure}

Channel maps of the CO data are shown in Figure \ref{co-chan}.

\section{Results}

\begin{figure}
 \epsscale{1.0}
 \includegraphics{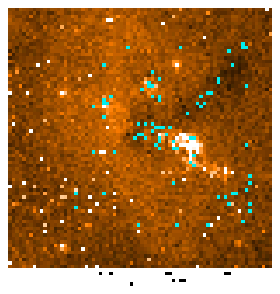}
 \vspace*{1.25in}
 \caption{\sii\ image with contours of 850\mum\ emission overlayed.  Contour
intervals same as in Figure \ref{850um}.
\label{sii-850um}} 
\end{figure}

The dust continuum image shown in Figure \ref{850um} clearly shows
three distinct regions of emission.  In Figure \ref{sii-850um} these
regions are identified with optical features, i.e. toward the HL~Tau
group, associated with IRS5 and NE and their outflows, and toward a dark
ridge or core extending north-west of IRS5.

All known Class 0/I YSOs (IRS5/NE/HL~Tau) were detected.  Two cTTs
(Class II) (XZ~Tau/LkH$\alpha$358) were detected near HL~Tau where the
RMS of the image is especially low.  No other cTTs or wTTs
(Class II/III) were detected at the RMS of the image.  No previously
unknown stellar
objects were detected.  We have detected one starless core, in which
Swift, Welch \& Di Francesco (2005) have found evidence for collapse.

The flux densities and integrated intensities of the detected sources
and regions are tabulated in Table \ref{tbl-1}.  These, as well as
deconvolved source sizes (from gaussian fits), were determined
using IDL 
routines.  Background emission was subtracted from all of the
compact or slightly resolved sources, while the derived flux densities
of the extended sources have had internal ``point'' sources already
subtracted.  Care should be taken, however, in interpreting the
results of these extended structures, since the maps were made by
chopping on the sky, and the image reconstruction method has difficulty
restoring large-scale structure (Johnstone \etal\ 2000).  

The mass of each source/region, shown in Table \ref{tbl-1}, was derived by
assuming a dust temperature of 20 K and an opacity law with
$\kappa_{850}$ = 0.02 cm$^2$ g$^{-1}$ at 850 \mum\  (implicit in this is the
assumption of a gas to dust mass ratio of 100). At a distance of
168 pc (Bertout, Robichon \& Arenou 1999) the integrated flux 
(S$_{850}$) to mass ratio is

$M_{clump} = 0.14 * \left(\frac{[exp(17/T_d) - 1]}{[exp(17/20) - 1]}\right)\
\left(\frac {\kappa_{850}} {0.02\,{\rm cm}^2\,{\rm g}^{-1}} \right)^{-1}\
\left(\frac{D_{L1551}}{(168 {\rm pc})}\right)^2  S_{850}$ M$_{\odot}$.

We shall discuss each region separately.

\subsection{HL~Tau Group}

\begin{figure}
 \epsscale{1.0}
 \includegraphics{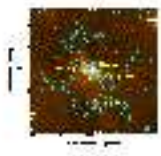}
 \vspace*{1.25in}
 \caption{Same as Figure \ref{sii-850um}, but focusing on the HL~Tau group.  
Contours are 0.015, 0.03, 0.06, 0.12, 0.24, 0.48, 0.96, and 1.92 
Jy beam$^{-1}$.
\label{hltau-sii-850}} 
\end{figure}

Figure \ref{hltau-sii-850} shows a close-up of the HL~Tau region in
\sii\ with
850\mum\  contours superimposed.  The \sii\ image shows the jets associated
with HL~Tau, HH~30* and XZ~Tau.  HL~Tau, a Class I protostar, is by far
the brightest source in this region, with peak intensity 2.2 and 9.6
Jy beam$^{-1}$ at 850 and 450 \mum\  respectively (Table
\ref{tbl-1}).  HL~Tau may have been barely resolved ($\sim$6\arcsec\ 
at 850\mum, $\sim$4.5\arcsec\ at 450\mum, fitted using IDL routines).

\begin{figure}
 \epsscale{1.0}
 \includegraphics{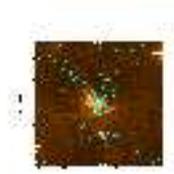}
 \vspace*{1.25in}
 \caption{\sii\ image of the HL~Tau region with 450\mum\ emission
contours overlaid.  Contour levels 
are 0.1, 0.2, 0.4, 0.8, 1.6 and 3.2 Jy beam$^{-1}$.  The ``+'' marks
the location of the proposed source VLA1 (Mundt, Brugel \& Buhrke  
1987).
\label{hltau-sii-450um}} 
\end{figure}

The deeply extincted edge-on disk source HH~30* was also clearly
detected at 0.05 Jy beam$^{-1}$ (Table \ref{tbl-1}).  At 450\mum\
(Figure \ref{450um}) it can be seen, but is difficult to extract from
the extended (plateau) emission.  The HH~30 jet does not seem to have any
related dust continuum emission.  LkH$\alpha$ 358 and XZ~Tau, two
Class II sources, were less clearly detected because of their
proximity to HL~Tau, but create significant ``deviation'' in the 850
and 450 \mum\ contours (Figures \ref{hltau-sii-850} \&
\ref{hltau-sii-450um}) around HL~Tau.    

Although we have clearly detected HL/XZ~Tau, LkH$\alpha$ 358 and
HH~30*, we do not see any evidence for a source VLA1-HL~Tau
(Brown, Drake \& Mundt 1985), located $\sim$12\arcsec\ east-north-east
of HL~Tau (see 
Fig. \ref{hltau-sii-450um}) and proposed to be the origin of a jet in the
vicinity of HL~Tau (Mundt, Brugel \& Buhrke 1987; Mundt, Buhrke \& Ray
1988).  This source has also not been
detected at any other wavelength, placing further doubt that VLA1 is a
young stellar object or protostar. 

Surrounding the stellar group is a broad ($\sim$2.5\arcmin$\times$4\arcmin),
``plateau'' of emission, with typical intensity $\sim$0.03-0.06 Jy beam$^{-1}$ and
integrated intensity $\sim$5 Jy after subtracting the point sources
(above).  This extended emission may be of concern for JCMT
calibration, since HL~Tau is one of the secondary flux calibrators for
SCUBA.

The HL~Tau jet at position angle $PA$ $\sim$ 51\arcdeg\
(Mundt \etal\ 1990; L\'{o}pez \etal\ 1995)  may drive HH~266 
located northeast
of these stars (e.g. Figures 1 \& 2 in Devine, Reipurth, \& Bally 1999). 
A 48\arcsec\ long blueshifted jet with a heliocentric radial
velocity ($v_{hel}$) = $-$180~\kms\ extends to the northeast and
a 65\arcsec\ long redshifted jet with $v_{hel}$ = +120~\kms\
extends to the southwest. 

An optical outflow originating from XZ~Tau can be traced more 
than 1\arcmin\ at $PA$ $\sim$ 15\arcdeg\ from this binary star. 
This flow is blueshifted with $v_{hel}$ = $-$45~\kms\ 
towards the northeast and redshifted with $v_{hel}$ = +77~\kms\ 
toward the southwest (see Krist et al. 1999 for HST images).   

The HH~30* jet at $PA$ $\sim$ 31\arcdeg\ lies within a couple
degrees of the plane of the sky (Mundt \etal\ 1990;
L\'{o}pez \etal\ 1995).  Mundt et al. (1990) traced this jet for
135\arcsec\ towards the northeast and 85\arcsec\ towards the 
southwest with a heliocentric velocity of $v_{rad}$ = +16 \kms\ 
towards both lobes.  On deep \Ha\ and \sii\ images, HH~30* can be 
traced to the main L1551 outflow lobe and possibly beyond.  Thus, 
this outflow may influence the HH complex located SW of IRS~5. 

At velocities between 8 \kms\ and 14 \kms\ (Figure
\ref{co-chan}) an arc of molecular gas extends from
the IRS~5 region towards the north. Lower resolution maps 
in \Tcoa\ and \tcoa\ led to the suggestion that this feature 
is related to the HH~30* outflow (Pound \& Bally 1991), but the 
\Tcob\ maps indicate that this feature lies east of HH~30*.  
It may, however, be related to one of the flows emerging from 
the HL-Tau complex.

\begin{figure}
\begin{center}
 \includegraphics{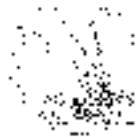}
 \vspace*{1.25in}
\caption{
  HL/XZ~Tau outflow in \Tcob. The blue lobe (solid contours) 
  is velocity integrated from 1.0 \kms\ to 6.0 \kms. The red 
  lobe (dotted contours) is velocity integrated from 9.0 
  \kms\ to 14 \kms. Contour levels = 3, 4, 5, ... K~\kms.
\label{hltau-co-jets}}
\end{center}
\end{figure}

The jet trajectories, overlaid on the \Tcob\ emission from
\vlsr\ = 2~\kms\ to 6~\kms and from \vlsr\ = 9~\kms\ to 14~\kms\, are
shown in Figure 
\ref{hltau-co-jets}. The blue-shifted CO emission near HL and XZ~Tau
forms a shell-like frame around the axes of the HL and XZ~Tau jets.
The CO has a
projected angular extent of about 5\arcmin\
or 0.2~pc and can be traced from \vlsr\ = 0.5~\kms\ to 
6.5~\kms\ where the lobe is lost in the ambient cloud.  There is no
apparent red-shifted emission associated with the southward jets,
although a finger of red-shifted emission may be tracing the northward
jet of HH~30*, which is nearly in the plane of the sky.

\subsection{L1551-IRS5/NE}

Figure \ref{irs5-sii-850} shows an optical \sii\ image of the region
near IRS5/NE and toward the south-west outflow lobe, with 850\mum\
dust emission contours superposed. Figure \ref{ne-sii-850} is a
similar figure toward the north-east of NE.  IRS5 is
located at the apex of an optical jet, while NE is not seen at all
optically.  NE has been classified as a Class 0 source
(Moriarty-Schieven, Butner \& Wannier 1995)
and IRS5 a Class I FU Ori protostar (Sandell \& Weintraub 2001).  

IRS5 is the brightest submillimeter source in the Taurus
complex, with peak intensity 3.16 Jy beam$^{-1}$.  NE has a peak
intensity of 1.16 Jy beam$^{-1}$ (Table \ref{tbl-1}).  Both are
extended ($\sim$10\arcsec), with large envelopes, that join with a
ridge of emission (Plateau in Table \ref{tbl-1}).   

Three known molecular outflows originate from this region, the IRS5
flow, the NE flow, and the L1551 E-W flow.  Optical imaging
(e.g. Devine, Reipurth, \& Bally 1999) indicates that some of these
flows may in turn be 
powered by several distinct sources.

The main IRS5 flow is oriented northeast-southwest, is
associated with HH~28, 29, 102, 258, 259, 262 \& 286, and may be
driven collectively by the IRS~5 binary and L1551~NE.
The recently recognized {\it L1551 NE Flow} associated with
HH~454 has the same orientation and is confused with
the main L1551 flow.  The {\it E-W flow} consists of a 
25\arcmin\ long redshifted east--west oriented CO ridge 
extending from HH~102 in the IRS~5 region towards the west.  

We look at each outflow in turn.

\subsubsection{The Main L1551 Flow}

Figure \ref{Billawal.Fig11.ps} shows the main L1551 outflow
along with the mean of the orientations of the two IRS~5 jets 
(Fridlund \& Liseau 1998) at $PA$ = 246\arcdeg.  The projected extent 
of the outflow is about 0.45\arcdeg\ or about 1~pc.  The southwest
lobe is blueshifted while the northeast lobe is redshifted 
with respect to the ambient cloud (centered at \vlsr\ = 6.75~\kms).  
The southwest lobe extends from $-$7~\kms\ to $\sim$ 8.6~\kms\ and consists of 
a hollowed out half-shell at low velocities that narrows at high \vlsr .  
The northeast lobe extends from \vlsr\ = 4.5~\kms\ to 22~\kms\ and is 
considerably smaller than the southwest lobe; it contains three connected clumps 
that surround a low-emission region with the clump furthest 
from IRS~5 coinciding with HH~262.  Unlike the southwest lobe, the northeast
lobe appears closed both spatially and kinematically.  Figure
\ref{Billawal.Fig12.ps} shows a velocity slice along a line
connecting HH~262, IRS~5, and HH~28.  Multiple peaks of
emission, both blueshifted and redshifted,  provide evidence
that both the southwest and northeast lobes have been affected by multiple
outbursts from a single source, or by separate outflows from
different sources. 

It is important to point out that we detected several HH objects 
associated with the IRS5/NE main outflow complex well beyond
the projected edge of the L1551 cloud (e.g. HH~286; Devine, Reipurth, 
Bally 1999).  HH~286 is a large but dim H$\alpha$ bow shock 20\arcmin\
northeast of IRS5.  Many galaxies are seen in this portion of the 
Mosaic images.  
The CO lobes only extend as far as the projected edge of the CO cloud.  
The outflow, however, continues on and produces shocks in either neutral 
or ionized gas. These shocks are seen as these HH objects beyond the 
projected edge of the cloud.

\begin{figure}
 \epsscale{1.0}
 \includegraphics{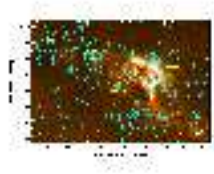}
 \vspace*{1.25in}
 \caption{Same as Figure \ref{sii-850um}, but focusing on the IRS5/NE
outflow.   
Contours are 0.02, 0.04, 0.08, 0.16, 0.32, 0.64, 1.28, and 2.56 
Jy beam$^{-1}$.
\label{irs5-sii-850}} 
\end{figure}

In Figure \ref{co-850} we plot contours of 850 \mum\ emission on
top of a grey-scale images of CO emission (10-12 and 4-6 km
s$^{-1}$).  In Figure \ref{co-850} (above) we see a ridge of
weak 850 \mum\ emission tracing the north limb of the blue-shifted
outflow from IRS5 out to HH~102.  This ridge can be seen more clearly
in Figure \ref{irs5-sii-850}.  HH~102 is also clearly detected at 850
\mum, coincident with the optical emission (Figure
\ref{irs5-sii-850} but offset to the west of the CO intensity peak
(Figure \ref{co-850}).  From the south edge of HH~102, a finger of
850 \mum\ emission extends eastward back toward IRS5, perhaps
indicating a bow shock.  In addition, faint extended emission can
be seen in Figure \ref{co-850} toward the peak of the CO emission
in the south limb of the IRS5 southeast lobe.

The continuum emission measured by SCUBA may be contaminated by strong 
molecular line emission within the passband (Johnstone \etal\ 2003).
Such conditions are most likely to occur for strong lines, such as
$^{12}$CO 3-2, in environments with broad kinematic features such as shocks
and jets.  We have analyzed the relatively few available $^{12}$CO 3-2 spectra
taken with the JCMT and determine that the typical integrated line strength
in the outflow lobes is $\sim$30 K km s$^{-1}$. 
This agrees well with the mapped
$^{12}$CO 2-1 observations presented in this paper. About one third of the
integrated line strength is due to the underlying molecular cloud and
varies only slightly across the region. This emission is chopped out
of the SCUBA observations. The remaining $\sim$20 K km s$^{-1}$ 
is highly variable
on scales of less than two arcminutes and may contribute to the observed
SCUBA flux. At this strength, the CO contamination is estimated to produce
$\sim$15 mJy beam$^{-1}$ of contaminated emission, compared with the typical
100 mJy beam$^{-1}$ measurement in the main L1551 lobe. Thus, contamination,
while not negligible, is not the main source of the observed SCUBA
emission, except possibly in the very weak continuum emission joining
IRS5 and HH~102, and toward the south limb of the southwest lobe. The
continuum emission, therefore, most likely comes from swept up dust
around the outflow lobe.

\subsubsection{The L1551~NE Flow}

\begin{figure}
 \epsscale{1.0}
 \includegraphics{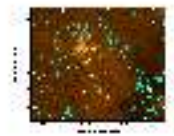}
 \vspace*{1.25in}
 \caption{Same as Fig. \ref{sii-850um}, but focusing on the region to the 
north-east of L1551 NE.  Contours same as 
Figure \ref{irs5-sii-850}.
\label{ne-sii-850}} 
\end{figure}

\begin{figure}
\begin{center}
 \includegraphics{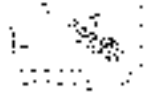}
 \vspace*{1.25in}
\caption{
  The main L1551 outflow in \Tcob. The southwest lobe (solid contours) is 
  velocity integrated from --6.0 \kms\ to 6.0 \kms. The NE lobe 
  (dotted contours) is velocity integrated from 7.0 \kms\ to 
 20.0 \kms. Contour levels = 16, 20, 24, ... K~\kms.
\label{Billawal.Fig11.ps}}
\end{center}
\end{figure}

\begin{figure}
\begin{center}
 \includegraphics{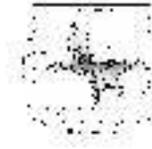}
 \vspace*{1.25in}
\caption{
  \Tcob\ velocity slice through HH~262, IRS~5, and HH~28. 
  Contours are for 0.4, 1.4, 2.4, ... K~\kms.
\label{Billawal.Fig12.ps}}
\end{center}
\end{figure}

\begin{figure}
 \epsscale{1.0}
 \includegraphics{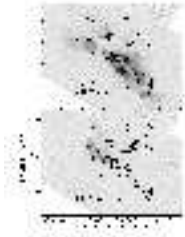}
 \vspace*{1.25in}
 \caption{
Contours of 850\,$\mu$m emission overlaid on a grey-scale image of the
blue-shifted CO emission (4-6\,km\,s$^{-1}$) (above), and red-shifted
CO emission (10-12\,km\,s$^{-1}$) (below).  Contour intervals the same
as in Figure \ref{850um}.
\label{co-850}} 
\end{figure}

Deep narrow-band images and proper motion measurements of 
HH~29 show that it is likely to be powered by a bipolar
HH flow from L1551~NE and {\it not} IRS~5 as had previously been thought 
(Devine \etal\ 2000).  HH~29 lies within a few degrees of the 
axis of a blueshifted jet visible in the 1.644 \mum\ line of [Fe II] 
that originates from L1551~NE and points towards the bow shock 
HH~29.   A chain of 
compact \sii\ emitting HH objects,  HH~454, lie along the axis of this jet 
(Figure \ref{Billawal.Fig11.ps}; Devine, Reipurth, \& Bally 1999).
Extending this jet axis through HH~454a (which can be seen in Figure
\ref{ne-sii-850} as a compact knot of \sii\ emission adjacent to NE to
the south-west) intersects HH~28 and HH~29.  Extending the axis to the
north-east intersects HH~454b/454c (seen in Figure \ref{ne-sii-850} as
slightly more diffuse knots $\sim$0.7\arcmin\ and $\sim$1\arcmin\ from
NE) and the weak bow shock of HH~262E at $04^h32^m10^s$
$+18{^{\circ}}12{\arcmin}15{\arcsec}$. 


In Figure \ref{ne-sii-850}, HH~262 has clearly been detected at 850 \mum.  
In addition, a weak ridge of emission appears to connect the north rim of 
HH~262 with NE, similar to the ridge of emission connecting HH~102 to
IRS5.  This emission, however, is of order 70 mJy beam$^{-1}$; though 
contamination due to CO emission may be significant at this intensity,
there is little 
apparent CO emission at this location
(Figure \ref{co-850}).

Given the roughly parallel jets from IRS5 and NE,
it is likely that the outflows from these sources are interacting with
each other.  Moriarty-Schieven, Butner \& Wannier (1995) first
detected the NE outflow, but because of limited mapping and confusion
with the IRS5 outflow they concluded that the blue-shifted outflow
lobe was oriented to the northeast, i.e. toward HH~262.  In Figure
\ref{co-850} (lower), however, the northeast lobe
red-shifted emission peaks at NE, and appears as a limb-brightened
ellipse with one end at NE, suggesting a significant amount of the
red-shifted outflow originates from NE.  The blue-shifted emission
within the northeast lobe (Figure \ref{co-850} upper) appears
disconnected from NE, and may be from the near side of the outflow
cavity.  There is significant blue-shifted emission peaking at NE, but
extending toward the southwest, as suggested by the fan-shaped
reflection nebula in this direction (Graham \& Heyer 1990).  This
suggests that the NE outflow is 
 mostly parallel to that from IRS5.  Indeed, Devine, Reipurth \& Bally
(1999) have proposed that HH~29 is driven by NE based on proper-motion
studies, and note that a nearly straight line can be drawn connecting
NE and HH~29 with HH~28, HH~262, and HH~286.

Devine, Reipurth \& Bally (1999) were
uncertain whether NE or IRS5 were the 
originating source of HH~262 based on their proper-motion studies,
although the orientation was suggestive.
Lopez et al (1998), also using proper
motion and kinematic 
studies, concluded that the proper motion of this source was
``roughly'' away from IRS5.  They dismissed the possibility that NE
was the origin, however, only because Moriarty-Schieven, Butner \&
Wannier (1995)  assumed the
NE blue lobe were oriented to the north-east toward HH~262, whereas
the kinematics of the HH object were red-shifted.  Our new CO data
confirm that the outflow orientation is the correct direction for NE
to be the origin of HH~262E, and, by extension, HH~286.  With HH~29 and
possibly HH~28 and HH~259 originating from NE, the south limb of the
blue-shifted outflow lobe may also originate from NE.  The ``hollow
shell'' morphology proposed for this outflow (Snell, Loren \& Plambeck
1980; Moriarty-Schieven \etal\ 1987,1988), may instead be due to two
interacting outflows.

\subsubsection{The E--W Outflow}

A third outflow originates from the region containing IRS5 and
L1551NE that extends due west from the L1551 cloud core
(Figure \ref{co-chan}, 8-12 km s$^{-1}$).  This feature, called the 
E--W flow (Moriarty-Schieven \& Wannier 1991; Pound \& Bally 1991),  
is a 30\arcmin\ 
(1.0~pc) long finger of redshifted gas at PA = 270\arcdeg\ 
extending from near IRS~5 toward the west with a maximum width of 
0.1~pc. The E--W flow has a relatively small velocity extent
ranging from about 6.5 \kms\ to 11.0 \kms\ and has no obvious 
blueshifted counterpart.  NE was originally proposed as the origin for
this outflow, but we see that this source powers an outflow parallel
to that from IRS5.
However, L1551~NE may be a multiple star system  
(Rodriguez, Anglada, \& Raga 1995; Moriarty-Schieven et al. 2000) 
and the companion star may be a viable candidate to drive the E--W flow.  

The new CO data shows that the eastern end of the E--W outflow lobe 
coincides with the northwestern rim of the main L1551 outflow shell 
and the Herbig-Haro object HH~102.  The lobe  
consists of a pair of nearly parallel filaments
of emission which delineate the walls of a 0.1~pc wide cavity. 
The filaments converge to pinch the cavity about
17\arcmin\ west of IRS~5 at the location of a relatively high
velocity and bright knot of CO (L1551~W; Pound \& Bally 1991).  
There is no known cloud core or IRAS source at this location.
The E--W flow continues west of this knot as a narrow and 
fading CO filament right to the boundary of the mapped region. 
Visual wavelength and near-infrared imaging have failed to 
reveal any emission line objects associated with 
the E--W outflow lobe. 

\begin{figure}
 \epsscale{1.0}
 \includegraphics{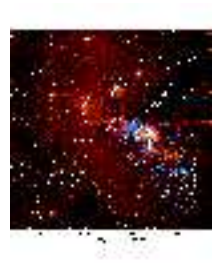}
 \vspace*{1.25in}
 \caption{
Comparison of the 850\,$\mu$m emission (white contours, same intervals
as Fig. \ref{850um}), CO outflows
(red-shifted (8-10 km s$^{-1}$) and blue-shifted (4-6 km s$^{-1}$)
emission as red and blue contours respectively (intervals at 2, 4, 6,
... K km s$^{-1}$)),, ambient CO emission (orange contour at 2 K km
s$^{-1}$), and optical \sii\ emission (greyscale) from the L1551 cloud. 
\label{sii-co-850}} 
\end{figure}

Figure \ref{sii-co-850} compares the 850
\mum, CO outflows, and optical emission from the cloud.  There is no
sign of HH objects, etc., within the outflow, or (especially) toward
the high-velocity peak ~20\arcmin west of 
IRS5.  Unfortunately the 850 \mum\ image doesn't extend far enough
west to tell if there is a continuum source at that position.  Thus
our data do not resolve the origin of this outflow.

\subsection{L1551-MC}

In figure \ref{sii}, a dark, heavily extincted bar can be seen
extending $\sim$16\arcmin\ north-west of IRS5.  About half way along this bar
lies a lone Herbig-Haro object, HH~265, whose origin is unknown.
We have detected 850 \mum\ emission associated with this
dark bar, in the neighborhood of HH~265.  The emission
(Fig. \ref{850um}) appears as a roughly elliptical, weakly centrally
condensed core, with peak intensity $\sim$0.2 Jy beam$^{-1}$,
and integrated intensity $\sim$4.5 Jy at 850 \mum.  

\begin{figure}
 \epsscale{1.0}
 \includegraphics{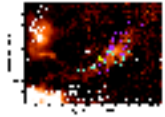}
 \vspace*{1.25in}
 \caption{850 \mum\ dust continuum image in greyscale, with overlaid
with contours of NH$_3$ (magenta, contours at intervals of 0.5, 1, 1.5, ... K km s$^{-1}$), CCS
(cyan, intervals 0.05, 0.1, 0.15, ... K km s$^{-1}$), CO J=2-1 wing emission (white, contours 1.25, 1.75, 2.25,... K km s$^{-1}$), integrated over 4-6 km s$^{-1}$, and {\sii}, in yellow, indicating the position and morphology of HH~265.
\label{mc-nh3-ccs-co}} 
\end{figure}

The 850 \mum\ continuum emission coincides with an NH$_3$ clump
discovered by Swift, Welch \& Di Francesco (2005) 
(Figure \ref{mc-nh3-ccs-co}), which they've dubbed L1551-MC.  
Their NH$_3$ core is roughly of size 2.5\arcmin$\times$1.11\arcmin\ 
(0.110$\times$0.054\,pc at 168\,pc distance) aligned
with the orientation of the dark bar, and has an average density of
10$^4$-10$^5$ cm$^{-1}$, kinetic temperature of T$_K \sim$ 9K, and
total mass of $\sim2M_{\odot}$ (Swift, Welch \& Di Francesco 2005).
If we, too, assume a dust temperature of 9K, then we derive a mass of
the clump of 2.6 M$_{\odot}$, which is very similar to their value.
(Using T$_d$=20K gives 0.6 M$_{\odot}$ (Table \ref{tbl-1}).)

The column density of dust emission peaks at the same location as
NH$_3$ emission peaks (Figure \ref{mc-nh3-ccs-co}).  
Swift, Welch \& Di Francesco (2005) found that
CCS emission, conversely, was depleted toward the column density
peak.  They found no evidence for an embedded protostar, but did find
that the kinematics of the CCS emission suggests that the outer
regions, at least, of the core are
undergoing infall.

If this core is indeed infalling and eventually collapses, will this
core form an isolated protostar or fragment into a small cluster of
stars similar to the HL/XZ~Tau group?  The total mass still within the
HL/XZ~Tau core is of order 1 M$_{\odot}$ or so, although we are likely
to have missed some emission because of chopping.  The mass of stars
is also of order 1-2 M$_{\odot}$, so roughly 50\% of the mass has gone
into protostars.  If the L1551-MC core is similar, then $\sim$1
M$_{\odot}$ will form into protostars, enough mass to form several
low-mass stars.

There is no evidence from the NH$_3$ data (Figure \ref{mc-nh3-ccs-co},
magenta contours) that the clump has begun to fragment, at least at
the $\sim$30'' resolution of the observations.  The $\sim$14''
resolution 850$\mu$m data is lumpier, but the ``lumpiness'' is of the
same order as the variance in this part of the image, and thus shows
no evidence for fragmentation of the clump.

\subsection{Origin of HH~265}

\begin{figure}
 \epsscale{1.0}
 \includegraphics{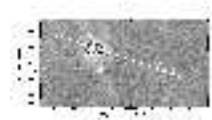}
 \vspace*{1.25in}
 \caption{Greyscale of the difference image between H$\alpha$ and
\sii\ toward the HL~Tau region, and showing the orientation of a
possible jet from LkH$\alpha$358 or HL~Tau, culminating at HH~265  The
greyscale is ``wrapped'' to display subtle features.  Knots along the
suggested jet toward HH~265 are indicated as a, b, and c.
\label{hh265-sii-halpha}} 
\end{figure}

The HH object HH~265 lies on the south-eastern end of the L1551-MC
core (Figure \ref{mc-nh3-ccs-co}).  It is unlikely that the origin of
this object lies within the clump, since the clump appears to be
prestellar.  Furthermore the morphology of the HH object is that of a
working surface ahead of a bow shock (Figure \ref{mc-nh3-ccs-co}),
with orientation of the complex extending north-eastward toward the HL~Tauri
group (Fig. \ref{hltau-sii-850}).  Furthermore, we see weak
blue-shifted CO emission within the 
bow shock, extending still further north-east in the direction of the HL~Tau
group, one member of which must therefore be the source of this HH
object.  None of the known jets (Figure \ref{hltau-co-jets}), however,
is oriented in the direction of HH~265, and all members of the group
except LkH$\alpha$358 are known to have jets.  Thus LkH$\alpha$358 is
a likely candidate for the origin of this HH object, although it must
be noted that a plume of emission associated with LkH$\alpha$358
(Fig. \ref{hltau-sii-850}) is oriented nearly due west, and not in the
direction of HH~265.  A line linking HH~265 with LkH$\alpha$358 (shown
in Fig. \ref{hh265-sii-halpha}) passes
very near to two faint H$\alpha$ emission knots at $4^h31^m19.8^s$
$+18{^{\circ}}12{\arcmin}30.3{\arcsec}$ (knot A in
Fig. \ref{hh265-sii-halpha}) and $4^h31^m23.2^s$ 
$+18{^{\circ}}12{\arcmin}54.5{\arcsec}$ (knot B).  Extending this line further
intersects another knot of H$\alpha$ emission at $4^h31^m49.4^s$
$+18{^{\circ}}14{\arcmin}50.7{\arcsec}$ (knot C).  This line also comes very
close to 
passing through HL~Tau, and comes within only a few arcsec of XZ~Tau.
If these emission features are
related, it would indicate a new, previously unknown jet originating
from LkH$\alpha$358, or possibly from the binary companion of XZ~Tau,
or a currently unkown binary companion of HL~Tau.

\section{Discussion and Summary}

In this paper, we present new sub-millimeter dust
continuum maps and J=2-1 CO images of the L1551 cloud
in the Taurus cloud complex.   These data are combined
with deep visual wavelength images and analyzed in the 
context of the extensive literature on this nearby 
star forming cloud.  

All previously known Class 0 and Class I YSOs in L1551 
are detected in the 850 $\mu$m sub-mm continuum images.
Furthermore, no new compact sub-mm sources were found.
Thus, the inventory of highly embedded young stellar
objects appears to be complete.  However, it is possible
that the known Class 0/I protostars contain additional members which
remain unresolved in our data.  

The compact sub-mm emitters are concentrated in two
sub-clusters.  The binary system IRS5, the most luminous 
and massive YSO in L1551,  and the L1551-NE multiple 
system, are located near the southeastern end of the 
cloud.  Both members of IRS5 and one member of the 
NE system power nearly parallel jets which appear to 
be collectively responsible for powering the main CO 
outflow in this cloud.  While the IRS5 binary and its 
HH~154 jet clearly power HH~102 and the northern rim of 
the CO outflow complex, NE is responsible for the highly 
collimated HH~454 jet and \sii\ knots, the bright HH 
object HH~29, possibly HH~28, the eastern portion of 
HH~262, and possibly the distant bow shock HH~286.   
Thus, one member of the NE system is responsible for 
the southern portion of the main CO outflow lobes.   

The driving source of the large redshifted east-west CO 
lobe originating near HH~102 remains unclear. Our 
SCUBA observations did not find any additional YSO 
candidates along the axis of this outflow lobe.  However, 
the orientation of this limb-brightened low-velocity outflow 
lobe suggests that the driving source may one member of 
the L1551-NE multiple star syst. 
   
The second sub-cluster of sub-mm emitters consists of
the sources clustered around HL~Tau (HH~30*, XZ~Tau,
and LkH$\alpha$~358).  All four of these YSOs are associated
with 850 $\mu$m emission with HL~Tau being by far the
brightest.  LkH$\alpha$~358 (or perhaps the binary companion of XZ~Tau
or a currently unknown binary companion of HL~Tau) may be the origin
of a jet terminating at HH~265.  A very weak molecular outflow was detected
associated with the end of this jet.

In addition to the compact sources, the sub-mm continuum
emission exhibits diffuse components.  Ridges of 850 
$\mu$m emission extends from IRS5 to HH~102, from IRS5
towards the axis of the southwestern CO lobe, from IRS5
towards NE, and from from NE towards HH~262.  The brightest
parts of both HH~102 and HH~262 are detected at 850 $\mu$m.
Although CO J = 3-2 emission may make a small contribution
to this emission, most of the flux is likely to be produced
by warm dust entrained by outflows.  A complex halo of dust
emission also surrounds the HL~Tau region.  This extended 
emission may compromise the use of this YSO as a secondary
calibrator for sub-mm continuum observations.

The L1551 cloud is cometary, indicating progressive erosion
from the general direction of Orion, located to the southeast.
Deep wide-field visual wavelength images and CO maps show
that the L1551 cloud has a cometary morphology consisting
of a dense head near the IRS5 / NE and HL~Tau region, and
a diffuse tail extending more than a degree towards the
northwest.  The densest part of this diffuse tail can
be clearly seen in our \sii\ image (Figure 4).  Diffuse
850 $\mu$m dust emission is produced by this region, dubbed
L1551-MC.  This object is likely to be a pre-stellar core showing
signs of undergoing the first phases of gravitational
collapse (Swift, Welch, \& Di Francesco 2005).  We do not see
any signs of fragmentation of the clump.  Assuming a 
dust temperature of 9 K, our sub-mm continuum flux implies a 
total mass of order 3 M$_{\odot}$.  

The wide-field \sii\ image (Figure 4) shows that the southeastern
rim of the L1551 cloud is illuminated from the southeast.  
The CO maps show that the cloud has a sharp edge in this direction.
Many background galaxies are visible beyond the southeastern edge
of the cloud, indicating that this portion of the cloud has been
dispersed.  The locations of the older Class II and Class III 
YSOs born from the L1551 cloud indicate that stars formed 
relatively recently southeast of the current cloud edge (Figure 4).  
The cometary shape of the L1551 cloud, the projected
distribution of older YSOs, the faint glow along the southeastern 
rim of L1551, and the location of L1551-MC northwest of the
recently formed Class 0/I YSOs indicate that the L1551 cloud 
has been irradiated and eroded from the general direction of Orion. 

The currently active sites of star formation in Orion are
located between 380 and 470 pc from the Sun.  However, the older
sub-groups of the Orion OB association are located closer with 
some members of the 1a and 1b subgroup as near as 300 pc, about
150 pc from L1551 (de Zeeuw et al 1999).  The closest portion
of the Orion-Eridanus supershell, the H$\alpha$ feature known 
as the Eridanus Loop (Reynolds, \& Ogden 1979; Boumis et al. 2001), 
has been estimated to be only 160 pc from the Sun  (Burrows \& Guo 
1996), which places it within 50 pc of L1551.  Portions of the 
21 cm HI supershell driven  by the Orion OB association and 
100 $\mu$m dust emission traced by the IRAS satellite are located 
well outside the H$\alpha$ loops illuminated by Orion's massive 
stars (e.g. Brown, Hartnamm, \& Burton 1995; Heiles, Haffner, 
\& Reynolds 1999).

The two well-known massive stars, Betelgeuse and Rigel, are located
in the closer-parts of the Orion-Eridanus supershell and southeast 
of L1551.  Betelgeuse, is about 130 pc from the Sun and has an 
18 \kms\ proper motion towards PA $\sim$ 68\arcdeg . 
Rigel is located 240 pc from the Sun and illuminates a 
small complex of star-forming molecular clouds associated
with IC 2118 in the southwester portion of Orion (Kun et al.
2004).  The proper motion of Betelgeuse (7.6 milliarcsec /year)
would place it southeast of L1551 several million years ago.
Thus, both of these massive stars could have been within 
100 pc of the L1551 cloud within the last several million years 
and may have contributed to the photo-erosion of this cloud.   

In summary, distribution of YSOs, the cometary shape, the
external illumination, and location of the pre-stellar core
L1551-MC indicate that the evolution of the L1551 cloud
has been influenced by massive stars and photo-erosion from
the general direction of Orion.  The closest parts of the
Orion OB associations, and massive stars in the direction of
Orion may be responsible for the evolution of the L1551 cloud.

\acknowledgements

The research of D.J.\ is supported through a grant from the
Natural Sciences and Engineering Research Council of Canada.  Thanks to 
Youssef Billawala for supplying the reduced CO J=2-1 map of the region and
to Jonathon Swift for supplying the molecular line maps of L1551-MC.
The JCMT is operated by the Joint
Astronomy Centre on behalf of the Particle Physics and Astronomy
Research Council of the UK, the Netherlands Organization for
Scientific Research, and the National Research Council of Canada.  The
authors acknowledge the data analysis facilities provided by the
Starlink Project which is run by CCLRC on behalf of PPARC. 
Data mining was done using facilities of the Canadian Astronomy Data
Centre, which is 
operated by the Dominion Astrophysical Observatory for the National
Research Council of Canada's Herzberg Institute of Astrophysics.

\begin{deluxetable}{lccrrrrr}
\tabletypesize{\footnotesize}
\tablecolumns{8}
\tablewidth{0pc}
\tablecaption{Submillimeter Continuum Properties of Sources in L1551\label{tbl-1}}
\tablehead{
\colhead{Source}   		 & 
   \colhead{Peak$_{850{\mu}m}$}  &
   \colhead{Integ$_{850{\mu}m}$} & 
   \colhead{{\tablenotemark{a}}Size$_{850{\mu}m}$}  & 
   \colhead{Peak$_{450{\mu}m}$}  & 
   \colhead{Integ$_{450{\mu}m}$} &
   \colhead{Size$_{450{\mu}m}$}  & 
   \colhead{Mass\tablenotemark{b}} \\
\colhead{} 			 & 
   \colhead{Jy beam$^{-1}$} 	 &
   \colhead{Jy} 		 & 
   \colhead{arcsec} 	 	 & 
   \colhead{Jy beam$^{-1}$} 	 & 
   \colhead{Jy} 		 &
   \colhead{arcsec} 		 & 
   \colhead{M$_{\odot}$}
}
\startdata
{\em HL~Tau Region} \\
HL~Tau & 2.15 & 2.29 & 6.3$\times$6.0 & 9.36 & 9.62 & 4.5$\times$4.4 & 0.32 \\
LkH$\alpha$ 358 & 0.05 & \nodata & \nodata & $<$0.13 & \nodata & \nodata & 0.007 \\
XZ~Tau & 0.01 & \nodata & \nodata & $<$0.20 & \nodata & \nodata & 0.0014 \\
HH~30* & 0.05 & \nodata & \nodata & $<$0.13 & \nodata & \nodata & 0.007 \\
Plateau & 0.05 & 5.23 & 80.7$\times$51.8 & & & & 0.73 \\
\hline
{\em L1551 Region} \\
IRS5 & 3.16 & 7.23 & 10.2$\times$8.0 & 17.4 & 45.7 & 8.1$\times$6.2 & 1.01 \\
NE & 1.16 & 2.78 & 9.8$\times$8.8 & 5.7 & 8.33 & 7.4$\times$6.9 & 0.39 \\
Plateau & 0.18 & 19.05 & 99.0$\times$40.5 & & & & 2.67 \\
North-east Lobe & & $\sim$0.2 & & & $\sim$2.9 & & 0.03 \\
South-west Lobe & & $\sim$0.3 & & & \nodata & & 0.04 \\
\hline
{\em L1551-MC/ Dark Arc} \\
L1551-MC & 0.05 & 4.5 & 100$\times$33.2 & & & & 0.63 \\
Embedded Source & 0.04 & 0.2 & 16.1$\times$12.0 & & & & 0.028 \\
\enddata
\tablenotetext{a}{Deconvolved sizes.}
\tablenotetext{b}{Mass derived from the total flux at 850\mum\ assuming
$T_d = 20\,$K and $\kappa_{850} = 0.02\,$cm$^{2}$g$^{-1}$.}
\end{deluxetable}
                                                                                
\clearpage

\pagebreak

\clearpage

\end{document}